\documentstyle[prd,floats,aps,epsf,twocolumn]{revtex}
\def\Journal#1#2#3#4{{#1} {\bf #2}, #3 (#4)}

\def\NIM{Nucl. Instrum. Methods}

\def\NPB{{Nucl. Phys.} B}
\def\PLB{{Phys. Lett.}  B}
\def\PRL{Phys. Rev. Lett.}

\def\PRD{{Phys. Rev.} D}

\def\ZPC{{Z. Phys.} C}

\newcommand{\etal}{{\it et al.,}}
\newcommand{\Z}{Z^{0}}
\newcommand{\Met}{\mbox{$\raisebox{.3ex}{$\not$}E_{T} $}}

\begin{document}                                             
\begin{titlepage}
\begin{center}
\begin{bf}
{Search for New Heavy Particles in the $WZ^{0}$ Final State in $p{\bar p}$ Collisions at $\sqrt{s} = 1.8$ TeV}
\end{bf}

\end{center}
\begin{small}
\hfilneg


\font\eightit=cmti8
\def\r#1{\ignorespaces $^{#1}$}
\hfilneg
\begin{sloppypar}
\noindent
T.~Affolder,\r {23} H.~Akimoto,\r {45}
A.~Akopian,\r {37} M.~G.~Albrow,\r {11} P.~Amaral,\r 8
D.~Amidei,\r {25} K.~Anikeev,\r {24} J.~Antos,\r 1
G.~Apollinari,\r {11} T.~Arisawa,\r {45}
% ...
\end{sloppypar}
\vskip .026in
\begin{center}
(CDF Collaboration)
\end{center}

\vskip .026in
\begin{center}
\r 1  {\eightit Institute of Physics, Academia Sinica, Taipei, Taiwan 
11529,
Republic of China} \\
\r 2  {\eightit Argonne National Laboratory, Argonne, Illinois 60439} 
\\
\r 3  {\eightit Istituto Nazionale di Fisica Nucleare, University of
Bologna,
I-40127 Bologna, Italy} \\
\r 4  {\eightit Brandeis University, Waltham, Massachusetts 02254} \\
\r 5  {\eightit University of California at Davis, Davis, California  
95616}
\\
\r 6  {\eightit University of California at Los Angeles, Los
Angeles, California  90024} \\
\r 7  {\eightit Instituto de Fisica de Cantabria, CSIC-University of
Cantabria,
39005 Santander, Spain} \\
\r 8  {\eightit Enrico Fermi Institute, University of Chicago, Chicago,
Illinois 60637} \\
\r 9  {\eightit Joint Institute for Nuclear Research, RU-141980 Dubna,
Russia}
\\
\r {10} {\eightit Duke University, Durham, North Carolina  27708} \\
\r {11} {\eightit Fermi National Accelerator Laboratory, Batavia, 
Illinois
60510} \\
\r {12} {\eightit University of Florida, Gainesville, Florida  32611} 
\\
\r {13} {\eightit Laboratori Nazionali di Frascati, Istituto Nazionale 
di
Fisica
               Nucleare, I-00044 Frascati, Italy} \\
\r {14} {\eightit University of Geneva, CH-1211 Geneva 4, Switzerland} 
\\
\r {15} {\eightit Glasgow University, Glasgow G12 8QQ, United 
Kingdom}\\
\r {16} {\eightit Harvard University, Cambridge, Massachusetts 02138} 
\\
\r {17} {\eightit Hiroshima University, Higashi-Hiroshima 724, Japan} 
\\
\r {18} {\eightit University of Illinois, Urbana, Illinois 61801} \\
\r {19} {\eightit The Johns Hopkins University, Baltimore, Maryland 
21218}
\\
\r {20} {\eightit Institut f\"{u}r Experimentelle Kernphysik,
Universit\"{a}t Karlsruhe, 76128 Karlsruhe, Germany} \\
\r {21} {\eightit Center for High Energy Physics: Kyungpook National
University, Taegu 702-701; Seoul National University, Seoul 151-742; 
and
SungKyunKwan University, Suwon 440-746; Korea} \\
\r {22} {\eightit High Energy Accelerator Research Organization (KEK),
Tsukuba,
Ibaraki 305, Japan} \\
\r {23} {\eightit Ernest Orlando Lawrence Berkeley National Laboratory,
Berkeley, California 94720} \\
\r {24} {\eightit Massachusetts Institute of Technology, Cambridge,
Massachusetts  02139} \\
\r {25} {\eightit University of Michigan, Ann Arbor, Michigan 48109} \\
\r {26} {\eightit Michigan State University, East Lansing, Michigan  
48824}
\\
\r {27} {\eightit University of New Mexico, Albuquerque, New Mexico 
87131}
\\
\r {28} {\eightit The Ohio State University, Columbus, Ohio  43210} \\
\r {29} {\eightit Osaka City University, Osaka 588, Japan} \\
\r {30} {\eightit University of Oxford, Oxford OX1 3RH, United Kingdom} 
\\
\r {31} {\eightit Universita di Padova, Istituto Nazionale di Fisica
          Nucleare, Sezione di Padova, I-35131 Padova, Italy} \\
\r {32} {\eightit University of Pennsylvania, Philadelphia,
        Pennsylvania 19104} \\
\r {33} {\eightit Istituto Nazionale di Fisica Nucleare, University and
Scuola
               Normale Superiore of Pisa, I-56100 Pisa, Italy} \\
\r {34} {\eightit University of Pittsburgh, Pittsburgh, Pennsylvania 
15260}
\\
\r {35} {\eightit Purdue University, West Lafayette, Indiana 47907} \\
\r {36} {\eightit University of Rochester, Rochester, New York 14627} 
\\
\r {37} {\eightit Rockefeller University, New York, New York 10021} \\
\r {38} {\eightit Rutgers University, Piscataway, New Jersey 08855} \\
\r {39} {\eightit Texas A\&M University, College Station, Texas 77843} 
\\
\r {40} {\eightit Texas Tech University, Lubbock, Texas 79409} \\
\r {41} {\eightit Institute of Particle Physics, University of Toronto,
Toronto
M5S 1A7, Canada} \\
\r {42} {\eightit Istituto Nazionale di Fisica Nucleare, University of
Trieste/
Udine, Italy} \\
\r {43} {\eightit University of Tsukuba, Tsukuba, Ibaraki 305, Japan} 
\\
\r {44} {\eightit Tufts University, Medford, Massachusetts 02155} \\
\r {45} {\eightit Waseda University, Tokyo 169, Japan} \\
\r {46} {\eightit University of Wisconsin, Madison, Wisconsin 53706} \\
\r {47} {\eightit Yale University, New Haven, Connecticut 06520} \\
\r {(\ast)} {\eightit Now at Carnegie Mellon University, Pittsburgh,
Pennsylvania  15213} \\
\r {(\ast\ast)} {\eightit Now at Northwestern University, Evanston, 
Illinois
60208}
\end{center}

\end{small}
\vspace{1cm}
\begin{abstract}

\newcommand{\prlorigabstractlast}
{

We present a search for new heavy particles, $X$, which decay via $X \rightarrow W\Z \rightarrow e\nu +jj$  in $p{\bar p}$ collisions at $\sqrt{s} = 1.8$~TeV.

 Limits are set at the 95\% C.L. on the mass and the production of new heavy charged vector bosons which decay via $W'\rightarrow W\Z$ in extended gauge models as a function of the width, $\Gamma (W')$, and mixing factor between the $W'$ and the Standard Model $W$ bosons. 
} 

We present the first general search for new heavy particles, $X$, which decay via $X\rightarrow W\Z \rightarrow e\nu +jj$ as a function of $M_X$ and $\Gamma (X)$ in $p{\bar p}$ collisions at $\sqrt{s} = 1.8$~TeV.  No evidence is found for production of $X$ in 110 pb$^{-1}$ of data collected by the Collider Detector at Fermilab. General cross section limits are set at the 95\% C.L. as a function of mass and width of the new particle. The results are further interpreted as mass limits on the production of new heavy charged vector bosons which decay via $W'\rightarrow W\Z$ in an extended gauge model as a function of the width, $\Gamma (W')$, and mixing factor between the $W'$ and the Standard Model $W$ bosons.

\end{abstract}
\normalsize
\vspace*{.25in}
PACS numbers: XXX\\

\end{titlepage} 
\vspace{0.3cm}

\newcommand{\prlorigintro}{
The Standard Model (SM) of particle physics is widely believed to be incomplete. Because alternative models are numerous and varied, it is advantageous to search for new physics using methods that are not specific to a single model, but which retain the most compelling aspects of currently favored scenarios~\cite{Sleuth}. Searching for  new high mass particles, $X$, which decay via $X\rightarrow W\Z \rightarrow \ell \nu jj$ has the theoretical advantage that many models predict new particles with large couplings to gauge bosons, and the experimental advantages of a large branching ratio for $\Z \rightarrow jj$ and a striking signature of $W\rightarrow \ell \nu$.  One example of $X\rightarrow W\Z$, common in extended gauge models~\cite{Ramond}, is the production of a new heavy charged vector boson $W'$, where $W'\rightarrow W\Z$. In this case the width, $\Gamma (W')$, can vary greatly and is a function of the mixing factor, $\xi$, between the $W'$ and the $W$. Other models such as Technicolor with $\rho_T \rightarrow W\Z$~\cite{Technicolor} also exist. 
} 


The Standard Model (SM) of particle physics is widely believed to be incomplete. Because alternative models are numerous and varied, it is advantageous to search for new physics using methods that are not specific to a single model, but which retain the most compelling aspects of currently favored scenarios~\cite{Sleuth}. 

\newcommand{\checktexone}{\cite{Pati-Ramond},\cite{Non-linear},\cite{Technicolor}}
\newcommand{\checktexonereg}{\cite{Pati-Ramond,Non-linear,Technicolor}}

A number of theories, including extended gauge models, non-linear realizations of electroweak theory, a strongly interacting Higgs, and Technicolor, all predict new high mass particles $X$ which decay via $X \rightarrow W\Z$~\checktexonereg. A general $X \rightarrow W\Z$ search can address all of these models, as well as new theories that may be proposed in the future. While typical searches in $p{\bar p}$ collisions, such as for a Technicolor $\rho_T \rightarrow W\Z$~\cite{Technicolor}, consider the narrow resonance case, $\Gamma(X) \ll M_X$, there are good reasons to consider a general search which looks for $X$ both as a function of mass and the width, even for large widths. For instance, a new heavy charged vector boson, $W'$, has a width which can vary greatly as it depends on a mixing factor, $\xi$, between the $W'$ and the $W$~\cite{Altarelli}. 


%

\newcommand{\prlorigparatwo}
{In this Letter, we present a search for $X\rightarrow W\Z \rightarrow e\nu jj$ production as a 
as a function of $\Gamma (X)$ using $p{\bar p}$ collisions at $\sqrt{s} = 1.8$~TeV
using the Collider Detector at Fermilab (CDF). 
} 

In this Letter, we present the first general search for $X\rightarrow W\Z$ production as a function of $M_X$ and $\Gamma (X)$ using $p{\bar p}$ collisions at $\sqrt{s} = 1.8$~TeV using the Collider Detector at Fermilab (CDF). We use the final state $W\Z \rightarrow e\nu jj$ as it has the experimental advantages of a large branching ratio for $\Z \rightarrow jj$ and a striking signature of $W\rightarrow \ell \nu$.  
 The data, taken during the 1992--1995 Tevatron collider run, correspond to an integrated luminosity of
110 pb$^{-1}$.  Detailed descriptions of the
detector can be found elsewhere~\cite{NIM}. The portions of the detector
relevant to this search are: (i) a time projection chamber for vertex finding,  (ii) a drift chamber immersed in
a 1.4~T solenoidal magnetic field for tracking charged particles in the range
$|\eta|<1.1$~\cite{Pseudorapidity Ref}, and (iii) electromagnetic and hadronic calorimeters
covering the pseudorapidity range $|\eta|<4.2$. 
An electron is
identified
 as a narrow shower in the electromagnetic calorimeter that is matched in 
position
 with a track in the drift chamber.  Jets are reconstructed as clusters 
of energy in the calorimeter using a fixed-cone algorithm with cone 
size $\Delta R \equiv \sqrt{(\Delta\phi)^2 + (\Delta\eta)^2} = 0.4$.  The presence of neutrinos is inferred from the momentum
 imbalance, \Met, in the transverse plane as measured in the calorimeters.

 Candidate events are selected online using a three-level trigger system~\cite{NIM} 
to identify $W\rightarrow e\nu$ decays based on the requirement of an electron candidate with
 $E_T>22$~GeV, $|\eta|<1.1$ and a matching drift chamber track, and \Met\ $>22$~GeV.
 Several backup trigger paths, imposing for example electron $E_T>25$~GeV with no track requirement and \Met\ $>25$~GeV,  combine to make the trigger inefficiency for $X\rightarrow W\Z$ production negligible in 
the
 final $W\rightarrow e\nu$ sample.  To search for resonant $W\Z$ production, and to 
reduce
 standard model backgrounds, we raise the $E_T$ and \Met\ thresholds and
 require two jets to be present.  The final event selection requires 
an
 isolated electron~\cite{iso requirements} with $E_T>30$~GeV, \Met\ $>30$~GeV, 
and
 two jets with $E_T>50$~GeV and 20~GeV respectively, each with $|\eta|<2.0$.  
To
 reduce instrumental backgrounds, we restrict electrons to be in the fiducial region of the detector~\cite{fiducial}, and reject events in which significant 
hadron
 calorimeter energy is deposited out of time with the $p {\bar p}$ collision.  
A
 total of 512 events pass these requirements.

The acceptance, $A_{X}$, for the process $X \rightarrow W\Z \rightarrow e\nu jj$ is defined as the number of events originating from $X$ production and passing the final event selection, divided by the number of events in which $X\rightarrow W\Z$, $W\rightarrow e\nu$; the $\Z$ is allowed to have all decays. This definition allows non-quark decays of $\Z$, such as $\Z \rightarrow \tau^+ \tau^- \rightarrow jj$, to contribute to the acceptance. 
To compute $A_{X}$, we use the process $W' \rightarrow W\Z$  in the {\sc pythia} Monte Carlo (MC)~\cite{PYTHIA}, followed by a parametric simulation of the CDF detector.  We simulate $W' \rightarrow W\Z$ production for a variety of widths, $\Gamma (W')$. For narrow resonances, $\Gamma (W') \ll M_{W'}$, the acceptance rises from 7$\%$ at $ M_{W'} = 200$~GeV/c$^2$ 
to 31$\%$ at $M_{W'} = 600$~GeV/c$^2$. For a given mass, the acceptance falls with increasing particle width.


New production would show up as a resonance (peak) in both the dijet mass ($M_{\rm dijet}=M_{\Z}$) and the $W$+dijet mass ($M_{W+\rm dijet}=M_{X}$).  Since  a signal would appear as a clustered excess of events above the background spectrum, we search by analyzing the shape of the data in the dijet vs. $W$+dijet mass plane.  
Invariant masses are calculated using the measured energies and directions of the
electron, \Met, and the two jets.  To form the $W$+dijet mass we fix the mass of the electron+\Met\ system to be equal to the $W$ boson mass, which restricts the neutrino's unmeasured longitudinal momentum, $p_{z}^{\nu}$, to at most two possible values. When there are two solutions, we choose the one that yields a lower $W$+dijet invariant
mass. When there is no solution, we fix $p_{z}^{\nu}$ such that the reconstructed $W$ mass equals the transverse mass: $M_W = M_T \equiv {2p_{T}^{e} p_{T}^{\nu} [1 - \cos (\phi^{e} - \phi^{\nu})]}^{1/2}$. For $\Gamma (X) \ll M_X$, 
 MC studies show that on average these choices correctly reproduce the $\Z$ and $X$ masses with 15$\%$ resolution and no significant bias. For a given $\Gamma (X)$ the $W$+dijet mass distribution is given by this mass resolution and the intrinsic particle width.

The primary background to this search is SM $W$+jets production with
$W\rightarrow e\nu$. 
To estimate this background, we use the {\sc vecbos} MC~\cite{VECBOS} with $Q^2$=$<$$P^{\rm ~partons}_{T}$$>$$^2$+$M_W^2$, {\sc mrsdo$'$} structure functions~\cite{MRSD}, 
{\sc herwig} parton fragmentation~\cite{HERWIG}, and the detector simulation. 
The $W \rightarrow \tau \nu \rightarrow e\nu\nu\nu$ background is similarly
estimated but with {\sc tauola} MC~\cite{TAUOLA} used to model the decay $\tau \rightarrow e\nu\nu$.  We use {\sc vecbos} to model the kinematics of the events, but use the data for an overall normalization. 
Other backgrounds 
which produce the $e\nu jj$ final state include SM production of $t{\bar t}$, $W^+ W^-$, $t{\bar b}$, $W\Z$, $\Z (\rightarrow e^+e^- )$ + jets, 
$\Z (\rightarrow \tau^+\tau^- )$ + jets, and
multijet fakes. The $t{\bar t}$, $W^+ W^-$, $t{\bar b}$, and $W\Z$ backgrounds directly produce $e\nu jj$ events. Each is estimated using {\sc pythia} and the detector simulation, and
is normalized to the measured or theoretical cross sections~\cite{Background Cross Sections}.  We estimate that there are 45 $\pm$ 14 $t{\bar t}$ events, 9 $\pm$ 3 $W^+ W^-$ events, 3.0 $\pm$ 0.9 $t{\bar b}$ events and 1.6 $\pm$ 0.5
 $W\Z$ events in the data. 
The $\Z (\rightarrow e^+e^- )$ + jets and $\Z (\rightarrow \tau^+\tau^- )$ + jets
events can fake the $e$\Met $jj$ 
 signature if an electron  
from a
$\Z (\rightarrow e^+e^- )+\geq$ 2 jet event
is lost, faking the neutrino signature, or if in a $\Z +1$ jet event an energy
mismeasurement gives fake \Met\ and an electron or tau is misidentified as a jet.  We estimate the $\Z$+jets backgrounds using a combination of the
{\sc pythia} and {\sc vecbos} MC programs 
and the detector simulation, and normalize to the measured number of $\Z +1$ jet data events in the $\Z \rightarrow e^+e^-$ channel. We estimate that there are 36 $\pm$ 5 $\Z (\rightarrow e^+e^- )$ + jets events and 
1.6 $\pm$ 0.6 $\Z (\rightarrow \tau^+\tau^- )$ + jets events in the data.  
QCD multijet events can fake the $e\nu jj$ signature if a jet is misidentified as an electron and an energy mismeasurement in the calorimeter causes \Met. We estimate this background  from the data in a manner similar to 
that used in Ref.~\cite{W'->enu}, and predict that 
 27 $\pm$ 3 QCD multijet events remain in the final sample.
The contribution from all processes other than $W$+jets is 123 $\pm$ 16 events.

\begin{figure}[ntb]
\vspace*{-.45in}
\hspace*{.15in}
\mbox{\epsfysize=4in \epsffile{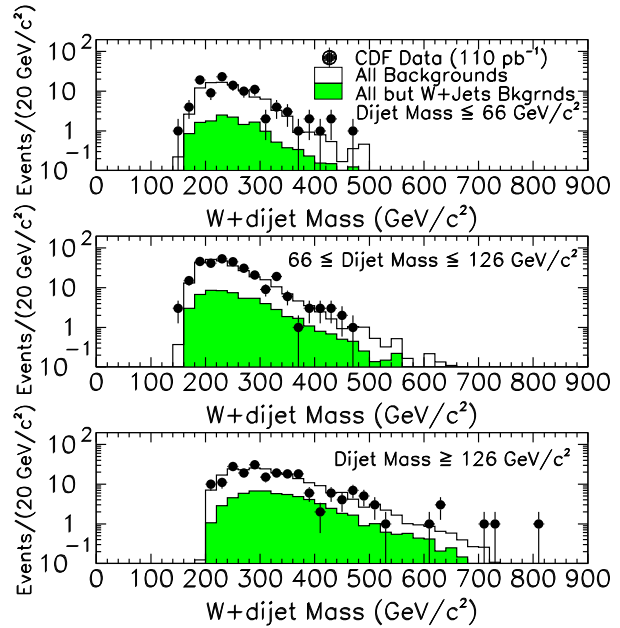}}
\hspace*{0.25in}
\vspace*{-.25in}
\caption{The $W$+dijet mass
spectra for the data and background 
with three different dijet mass requirements:
$ M_{\rm dijet} \leq 66$~GeV/c$^2$, $66\leq M_{\rm dijet}\leq 126$~GeV/c$^2$ and $M_{\rm dijet}\geq 126$~GeV/c$^2$.  A signal would appear as a resonance in the middle plot. The $W$+jets background spectrum is normalized as described in the text so that there are the same number of events in the data as in the backgrounds. The upper and lower plots show that the region outside the signal region is well modeled.} 
\label{WDijet Windows} 
\end{figure}

We use a binned likelihood fit in the dijet vs. $W$+dijet mass plane (20 GeV/c$^2$ $\times$ 20 GeV/c$^2$ bins) to search for resonant $W\Z $ production. All backgrounds except $W$+jets are normalized absolutely.  The normalization of the $W$+jets background in the fit  is fixed such that the sum of the signal and all backgrounds equals the number of events observed in the data. The relative magnitude of the signal and the $W$+jets background is the only free parameter in the fit~\cite{new_ref}.  The $W$+dijet mass spectrum for the data and background is shown in Fig.~\ref{WDijet Windows} for events with the dijet mass around $M_{\Z}$ and in the regions outside a 25~GeV/c$^2$ mass window, with the expected distributions plotted assuming no signal contribution.   
The results of the fit require no significant signal contribution and there is no evidence of resonant $W\Z$ production for any mass or width for the acceptance model.

\begin{figure}[ntb]
\vspace*{-.45in}
\hspace*{.15in}
\mbox{\epsfysize=4in \epsffile{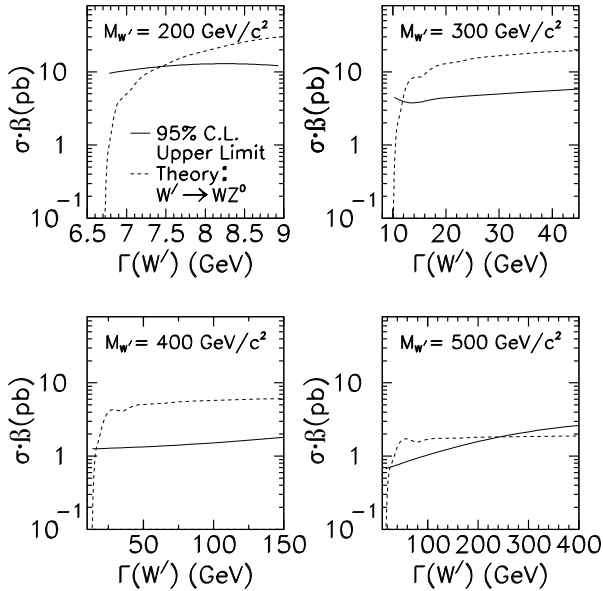}}
\hspace*{0.25in}
\vspace*{-.25in}
\caption{ The 95$\%$ C.L. upper limits on $\sigma \cdot {\cal B}$ as a function of the width. We use $\Gamma (W')$ as it uniquely determines the $W'\rightarrow W\Z$ branching ratio. Our results include the $W'\rightarrow W\Z$ and $W\rightarrow e\nu$ branching ratios.}
\label{Width Figure} 
\end{figure}

\newcommand{\prlorigsensativity}{
In order to examine the sensitivity to new particle production, we consider a class of models which extend the SM and contain heavier versions of the ordinary $W$ and $\Z$ vector bosons, labeled $W'$ and $Z'$~\cite{Altarelli}. 
In the simplest model there are no additional fermions, the $W'$ is a heavier version of the SM $W$, and the $W$ and $W'$ vertex couplings ($Wq{\bar q'}$, $W\ell \nu$ and
$WW\Z$) are identical.   
The dominant features of this reference model are the 
high production cross sections, and the increase in the partial width of the $W'$, 
$\Gamma (W'\rightarrow W\Z) \propto
 M_{W'}^5$, which yields a large branching fraction into $W\Z$.  
However, at large masses, $M_{W'}\approx 425$~GeV/c$^2$, $\Gamma (W')$ becomes so large that  
perturbation theory is no longer valid~\cite{Higgs Hunters Guide}.  While the presence of a direct $W'W\Z$ vertex makes the reference model itself implausible,  similar results can be obtained with a strongly interacting Higgs or non-linear realization of electroweak theory~\cite{Non-linear}.

Extended gauge models~\cite{Pati}, which restore left-right symmetry to the weak force, predict an effective $W'W\Z$ vertex term due to $W'$ and $W$ mixing. These models give the same $W'W\Z$ vertex as in the reference model, but multiplied by a mixing factor, $\xi$, which is estimated 
to be of the order $(\frac{M_W}{M_{W'}})^2$~\cite{Ramond}.  
Using $\xi = 1$ reproduces the reference model.  In this case the width only increases linearly with $M_{W'}$ and is small compared to both the mass and the detector resolution for all masses. 
Most previous searches for direct production and decay of $W'$ have assumed this type of result with small $\xi$ and searched in the $W' \rightarrow \ell \nu$ and $W' \rightarrow jj$
channels, establishing a limit of $M_{W'}>786$~GeV/c$^2$~\cite{W'->enu,Heavy W Bosons,split ref}. 
 However, if the $\nu$ is heavy, unstable, or highly interacting (as might be the case with right-handed neutrinos), or the $W' \rightarrow \ell \nu$ and $W' \rightarrow jj$ decays are otherwise forbidden, the branching fraction of $W' \rightarrow W\Z$ can be large. Although other models also predict new particles which have large branching fractions into $W\Z$, we set limits in the extended gauge model to allow quantitative comparisons.  Since $\xi$ directly affects the total width,  resonant production can be modeled as a function of either $\xi$ or $\Gamma (W')$. A previous search for $W'\rightarrow W\Z$ for $\Gamma(W') \ll M_{W'}$ can be found in~\cite{beat us out}.

} 

To set general limits on the process $p{\bar p} \rightarrow X \rightarrow W\Z$, we take $X$ to be a $W'$ in an extended gauge model as it spans both the $M_X$ and $\Gamma(X)$ parameter space. Following the prescription of Ref.~\cite{Altarelli} (no additional fermions and the $W$ and $W'$ vertex couplings, $Wq{\bar q'}$, $W\ell \nu$ and
$WW\Z$, are identical), the production cross sections are uniquely determined as a function of mass, and the partial width of the $W'$, 
$\Gamma (W'\rightarrow W\Z)$, is determined by a mixing factor, labeled $\xi$, which describes the amount of mixing between the $W$ and the $W'$.  While this makes $\Gamma$ a free parameter in the theory, we quote results in two specific cases. The full mixing case, or reference model~\cite{Altarelli}, is where the new particle $X$ couples in the same way as the SM $W$ ($\xi = 1$) and gives $\Gamma (W'\rightarrow W\Z) \propto M_{W'}^5$; yielding a large branching fraction into $W\Z$, and widths comparable to the mass for $M_{W'} \approx 425$~GeV/c$^2$. A second special case is $\xi = (\frac{M_W}{M_{W'}})^2$ as in extended gauge models which restore left-right symmetry to the weak force and predict an effective $W'W\Z$ vertex term~\cite{Pati-Ramond}. In this case, $\Gamma (W') \ll M_{W'}$ for all masses. 

 We set limits on $\sigma (p{\bar p} \rightarrow X) \cdot {\cal B}$, where ${\cal B}  = {\cal B} (X\rightarrow W\Z ) \cdot {\cal B}(W\rightarrow e\nu )$, using the fit technique described above and convoluting in systematic uncertainties, which depend on both mass and width, using the same methods as in Ref.~\cite{Heavy W Bosons}. The dominant source of uncertainty is the jet energy scale which would bias the measurement of the dijet and $W$+dijet masses from the new particle $X$. The effect of such a bias is largest at lower mass, where increased background in the signal region can cause a large variation in the cross section limit. For example, the effect is between 50$\%$ and 100$\%$ for the reference model and between 30$\%$ and 60$\%$ for $\xi = (\frac{M_W}{M_{W'}})^2$. 
Other notable sources of uncertainty are: 
uncertainty in the jet resolution (between 15$\%$ and 30$\%$), effect of the $Q^2$ scale on the $W$+jets background shape (between 5$\%$ and 
25$\%$), choice of parton distribution functions (between 10$\%$ and 30$\%$), uncertainty in $W'$ acceptance (between 5$\%$ and 30$\%$), and MC modeling of initial and final state radiation (between 5$\%$
and 15$\%$). 
  The total systematic  uncertainty
is found by adding the above sources in quadrature, and varies between 50-100\% for the reference model and 40-75\% for $\xi = (\frac{M_W}{M_{W'}})^2$.  

\newcommand{\checktextwo}{\cite{W'->enu},blah \cite{Heavy W Bosons},blah \cite{split ref}}
\newcommand{\checktextworeg}{\cite{W'->enu,Heavy W Bosons,split ref}}

The 95$\%$ C.L. upper limits on $\sigma\cdot {\cal B}$ for $M_X$=200, 300, 400 and 500~GeV/c$^2$ are shown in Fig.~\ref{Width Figure} as a function of the width. While these limits are not sensitive enough to set mass limits on $\rho_T \rightarrow W\Z$ production, they exclude a large region of $W'$ parameter space. Table~\ref{final results table} gives a summary of results for the $\Gamma (W') \ll M_{W'}$ approximation using $\xi = (\frac{M_W}{M_{W'}})^2$.  The results in Fig.~2 can be interpreted as the first cross section limits as a function of $W'$ width, and
Fig.~3 shows the first $W'$ exclusion region for $\xi$ vs. $M_{W'}$ where the theoretical cross section exceeds the calculated 95$\%$ C.L. upper limit. Other direct searches for $W'$ at the Tevatron in the $W' \rightarrow \ell \nu$ and $W' \rightarrow jj$ channels have established a limit of $M_{W'}>786$~GeV/c$^2$~\checktextworeg, but only in the region of $\xi \approx 0$ which is complementary to region excluded in Fig.~3, and are only valid for $\Gamma(W') \ll M_{W'}$. A previous search for $W'\rightarrow W\Z$~\cite{Quaero} sets cross section limits for $M_{W'}=200, 350$ and 500~GeV/c$^2$, but only for $\Gamma(W') \ll M_{W'}$. 

For a $W'$ in the reference model ($\xi =1$), we exclude the region $200\leq M_{W'}\leq 480$~GeV/c$^2$. For masses below 200~GeV/c$^2$, the widths are small and the reference model is excluded at the 95\% C.L. by the $W'\rightarrow \ell \nu$ results~\cite{Altarelli,W'->enu}. Since the reference model is no longer valid for masses above 425~GeV/c$^2$ ($\Gamma (W')$ becomes so large that perturbation theory is no longer valid~\cite{Higgs Hunters Guide}) the entire model is now excluded. These results are generally applicable to other new particles $X$ with wide widths~\cite{Non-linear}.

\begin{table}[ntb]
\caption {Final results from the $X\rightarrow W\Z$ search using 110 pb$^{-1}$ of data for $\Gamma (X) \ll M_{X}$. Here we have modeled the new particle production with $W'\rightarrow W\Z$ in an extended gauge model for the special case of $\xi = (\frac{M_W}{M_{W'}})^2$, where $\xi$ is the mixing factor between the $W'$ and the SM $W$ boson. Note that the 95$\%$ C.L. cross section upper limit from the fit is on $X\rightarrow W\Z$ with $W \rightarrow e\nu$. In the denominator of the acceptance, $A_X$, we have allowed $\Z$ to have all decays.}
\vspace*{.25in}
\begin{tabular}{c|c|c}
$M_{X}$    & $A _{X}$ & 95$\%$ C.L. $\sigma\cdot {\cal B}$ Limit\\
(GeV/c$^2$)&          & (pb)\\
\hline
200 & 0.07 & 9.5\\
300 & 0.17 & 4.5\\
400 & 0.24 & 1.3\\
500 & 0.29 & 0.7\\
600 & 0.31 & 0.5\\
\end{tabular}
\label{final results table}
\end{table}

\begin{figure}[ntb]
\vspace*{-.45in}
\hspace*{.25in}
\mbox{\epsfysize=4in \epsffile{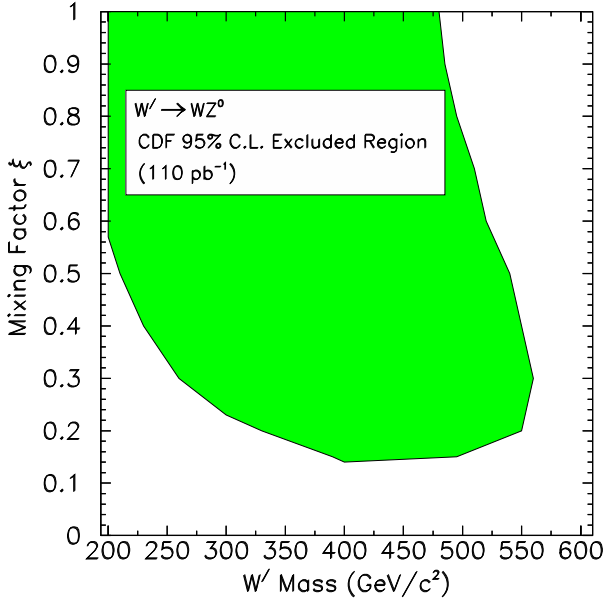}}
\vspace*{-.25in}
\epsfysize=4in
\caption{The 95$\%$ C.L. excluded region in the $\xi$ vs. $M_{W'}$ plane, where $\xi$ is the mixing factor between the $W'$ and the SM $W$ boson. While the branching fraction goes up with increasing values of $\xi$ and mass, the acceptance goes down as $\xi$ and the width increase. This causes the ``nose" effect in the exclusion region.  The largest mass exclusion occurs for $\xi$ = 0.3, where we exclude $M_{W'} < 560 $~GeV/c$^2$.} 
\label{Results Figures}
\vspace*{.1in}
\end{figure}

In conclusion, we have conducted a general search for new particles which decay via $X \rightarrow W\Z$ in the $e\nu jj$ channel. We observe 
 no evidence of resonant production and estimate production cross section limits as a function of mass and width. The results are further interpreted as mass limits on the production of new heavy charged vector bosons which decay via $W'\rightarrow W\Z$ in extended gauge models as a function of the width, $\Gamma (W')$, and mixing factor between the $W'$ and the $W$ bosons. These are the first limits on $X \rightarrow W\Z$ as a function of both mass and width, and are the only direct mass limits on $W' \rightarrow W\Z$ to date.



   We thank JoAnne Hewett, Tom Rizzo, and Jon Rosner for useful discussions and Jose Benlloch, Dan Hennessy, Marcus Hohlmann, Sacha Kopp, the Fermilab staff and the technical staffs of the
participating institutions for their vital contributions.  This work 
was
supported by the U.S. Department of Energy and National Science 
Foundation;
the Italian Istituto Nazionale di Fisica Nucleare; the Ministry of 
Education,
Science, Sports and Culture of Japan; the Natural Sciences and 
Engineering 
Research Council of Canada; the National Science Council of the 
Republic of 
China; the Swiss National Science Foundation; the A. P. Sloan 
Foundation; the
Bundesministerium fuer Bildung und Forschung, Germany; the Korea 
Science 
and Engineering Foundation (KoSEF); the Korea Research Foundation; and 
the 
Comision Interministerial de Ciencia y Tecnologia, Spain.


\begin{thebibliography}{99}

  
 
    \bibitem{Sleuth} Examples of such analyses include CDF Collaboration, F. Abe \etal\ \Journal{\PRL}{81}{1791}{1998};  D$\O$ Collaboration, B. Abbott \etal\ \Journal{\PRL}{86}{3712}{2001}. 

    \bibitem{Pati-Ramond} J.~Pati and A.~Salam, \Journal{\PRD}{10}{275}{1974};  
R.N.~Mohapatra and J.~Pati, \Journal{\PRD}{11}{566}{1975}; {\bf 11}, 2558 (1975); G.~Senjanovic and R.N.~Mohapatra, \Journal{\PRD}{12}{1502}{1975}; P.~Ramond, {\rm Ann. Rev. Nucl. Part. Sci.} {\bf 33}, 31 (1983). 



    \bibitem{Non-linear} P.~Chiappetta and S.~Narison, \Journal{\PLB}{198}{421}{1987}; R.~Casalbuoni  \etal\ \Journal{\NPB}{310}{181}{1988}.



    \bibitem{Technicolor}  K.~Lane and E.~Eichten, \Journal{\PLB}{222}{274}{1989}; K.~Lane, \Journal{\PRD}{60}{075007}{1999}; also see K.~Lane, hep-ph/0007304 (2000).

    \bibitem{Altarelli} G.~Altarelli, B.~Mele, and M.~Ruiz-Altaba, 
\Journal{\ZPC}{45}{109}{1989}. 
  
    \bibitem{NIM} CDF Collaboration, F.~Abe  \etal\ \Journal{\NIM}{271}{387}{1988}.  

    \bibitem{Pseudorapidity Ref} We use cylindrical coordinates where positive $z$ points along the proton beam 
and is zero at the center of the detector. The pseudorapidity, $\eta$, is defined 
as $\eta \equiv -\ln[{\rm tan}(\theta/2)]$, where $\theta$ is the polar angle with respect 
to the proton beam direction and $\phi$ is the azimuthal angle.  The transverse energy is defined as $E_T = E\sin\theta$, where $E$ is measured in the calorimeter.
   
    \bibitem{iso requirements} We require that the electron candidate pass identification and isolation requirements. The scalar sum of the $p_T$ of all tracks in the tracking chamber within a cone of $\Delta R = 0.25$ surrounding, but not including, the electron be less than 5~GeV/c, 
and that the $E_T$ in a cone of $\Delta R = 0.4$
around, but not including, the electron candidate be less than 10$\%$ of the electron $E_T$.
  

    \bibitem{fiducial} The fiducial region is $0.05\leq |\eta| < 1.05$ and away from the edges of the calorimeter; for a more complete discussion of the fiducial region, see section 2.c. of F.~Abe  \etal\ \Journal{\PRD}{52}{2624}{1995}.


    
    \bibitem{PYTHIA}
H.~Bengtsson and T.~Sj{\"o}strand, {\rm Comput. Phys. Commun.} {\bf 46}, 43 (1987).
 
    \bibitem{VECBOS} F.A.~Berends, W.T.~Giele, H.~Kuijf, and B.~Tausk,  \Journal{\NPB}{357}{32}{1991}.
 
    \bibitem{MRSD} A.D.~Martin, W.J.~Stirling, and R.G.~Roberts
\Journal{\PRD}{50}{6734}{1994}.
  
    \bibitem{HERWIG} We have interfaced the subsequent parton shower evolution and hadronization by the interface using the routines of G.~Marchesini and B.R.~Webber, \Journal{\NPB}{310}{461}{1988}; G.~Marchesini \etal\ {\rm Comput. Phys. Commun.} {\bf 67}, 465 (1992).
 

    \bibitem{TAUOLA} S.~Jadach, Z.~Was, R.~Decker, and J.H.~Kuhn, {\sc tauola} version 2.4, {\rm Comput. Phys. Commun.} {\bf 76}, 361 (1993). 

    \bibitem{Background Cross Sections} CDF Collaboration, F.~Abe  \etal\ \Journal{\PRL}{73}{2667}{1994}; S.S.D.~Willenbrock and  D.A.~Dicus, \Journal{\PRD}{34}{155}{1986}; 
J.~Ohnemus, \Journal{\PRD}{44}{1403}{1991}; {\bf 44}, 3477 (1991).
    
	\bibitem{W'->enu} CDF Collaboration, F.~Abe  \etal\ \Journal{\PRL}{74}{2900}{1995}. 
 
     \bibitem{new_ref} We note that the assumption of fixing the normalization of the non-$W$ backgrounds has between a 1\%-5\% effect on the limit.  This variation has been incorporated into the overall systematic uncertainty.


  

    
  

    \bibitem{Heavy W Bosons} CDF Collaboration, F.~Abe  \etal\ \Journal{\PRL}{74}{3538}{1995}. 

    \bibitem {split ref} CDF Collaboration, T. Affolder  \etal\ submitted to Phys. Rev. Lett., hep-ex/0107008; D$\O$ Collaboration,  S.~Abachi  \etal\ \Journal{\PLB}{358}{405}{1995}; \Journal{\PRL}{76}{3271}{1996}.
    
    \bibitem{Quaero} D$\O$ Collaboration, V.~M.~Abazov \etal\ \Journal{\PRL}{87}{231801}{2001}.

    \bibitem{Higgs Hunters Guide} J.~Gunion, H.~Haber, G.~Kane, and S.~Dawson,
{\it The Higgs Hunter's Guide}, Frontiers in Physics series (1989). 


    



   

 
  
\end{thebibliography}
\end{document}